\documentclass[amsmath,amssymb,altaffilletter,10pt,aps,pra]{revtex4-1}
\usepackage{graphicx}

\begin{document}

\title{Intensity autocorrelation measurements of quantum cascade laser frequency combs in the Terahertz range}%

\author{Ileana-Cristina  Benea-Chelmus}%
\email{ileanab@ethz.ch}
\author{Markus R{\"o}sch}
\author{Giacomo Scalari}
\author{Mattias Beck}
\author{J\'er\^ome Faist}
\email{jerome.faist@phys.ethz.ch}
\affiliation{ETH Zurich, Institute of Quantum Electronics, Auguste-Piccard-Hof 1, Zurich 8093, Switzerland}
\homepage[]{www.qoe.ethz.ch}

\date{\today}

\begin{abstract}
We report on the first direct measurement of the emission character of quantum cascade laser based frequency combs,  using intensity autocorrelation. The correlation technique is based on fast electro-optic sampling, with a bandwidth optimized to match the emission spectra of the comb laser, centered around 2.5~THz. We find the output light to be amplitude and frequency modulated at the same time, with intensity modulation depth as high as 90\%, occuring on timescales as short as few picoseconds. We compare these findings with intensity autocorrelation measurements of pulsed Terahertz light originating from a photoconductive antenna and find very different results. We observe no significant difference for the comb and dispersed regime. The technique will be of significant importance in future for the measurement of ultra-short pulses from quantum cascade lasers.
\end{abstract}


\maketitle

\section{Introduction}
Frequency combs~\cite{Delhaye:2007p1571} are sources of electromagnetic radiation, which have the intriguing property that their frequency components are equidistantly placed in frequency domain. This property qualifies them as highly precise spectral rulers, a merit which has promoted their use for high-resolution metrology~\cite{Udem:1999vw}, in the telecommunication domain~\cite{Pfeifle:2014cm}, as well as for spectroscopy~\cite{Holzwarth:2000p2058,Villares:2014gl,Roesch2016Dual}.  The traditional way to realize frequency combs was from mode-locked lasers~\cite{Udem:2002p2021}, and their output ideally represented a Fourier-limited train of pulses~\cite{Cundiff:2003um}. Nowadays, numerous alternative means  have been demonstrated, with notable results based on phase-locked lasers in the optical and near infrared~(NIR) domain~\cite{Jones:2000p2050}, Kerr-effect in microresonators for the NIR~\cite{Delhaye:2007p1571} and mid infrared~(MIR)~\cite{Griffith:2015gd,Wang:2013db}.

 In the MIR~\cite{Hugi:2012ep} and the Terahertz~(THz) region of the electromagnetic spectrum~\cite{Rosch:2014ft,Burghoff:2014hpa}, up to date the most compact and efficient frequency combs are from electrically injected quantum cascade lasers~(QCL)~\cite{Faist2016}. They have been shown to produce high-power, high bandwidth frequency combs, with Shawlow-Townes limited spectral purity~\cite{Bartalini:2010p1511,Cappelli2015}. The frequency comb is formed by Four Wave Mixing~\cite{Khurgin:2014hy}, a $\chi^{(3)}$ nonlinear process, where, in the frequency domain, the non-linear mixing of equidistant frequency components, is parametrically enhanced by energy conservation. In the time-domain, the underlying dynamics differ significantly from traditional frequency combs, being dominated by the very short upper state lifetime of intersubband transitions in comparison to the cavity roundtrip time~\cite{Faist:2013td}.

\begin{figure*}[hbtp]
  \centering 
  \includegraphics[width=\textwidth]{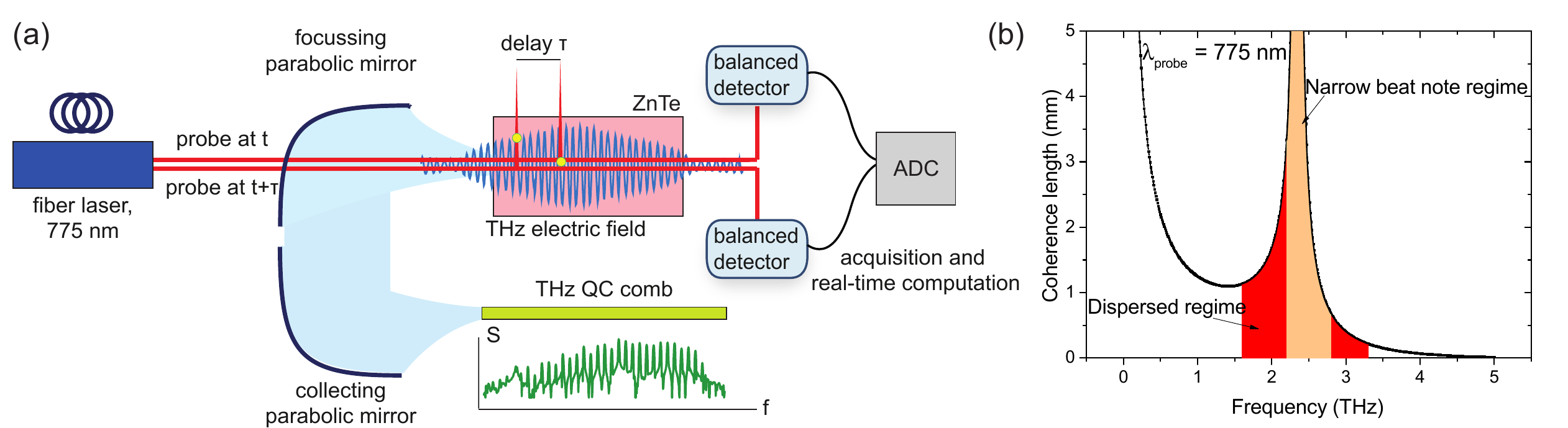}
  \caption{Measurement setup and detectivity bandwidth. (a) Time-domain correlation setup based on two-probe electro-optic sampling. The two probes sample the THz radiation of a QCL based frequency comb at two mutually delayed points in time. The probe time-constant is 146~fs. A fast acquisition module records the two sampled values and a further real-time routine computes from here the instantaneous intensity, and the two-point electric field and intensity product. When averaged over time and normalized this gives the coherence functions $g^{(1)}(\tau)$ and $g^{(2)}(\tau)$.b) Calculated coherence length of THz detection in (110)-oriented ZnTe, and a probe center wavelength of 775~nm, by using the formula $l_c=\frac{\pi c}{2\pi f_{THz} |n_{g,opt}-n_{ph,THz}|}$. The detectivity is matched to the laser comb emission, which is marked for when the investigated laser is operating in narrow beat note regime, 2.2~-~2.8~THz, (yellow) and dispersed regime, 1.6~-~3.3~THz (red).}
  \label{fig:setup}
\end{figure*}

Recently reported theroretical work attests that passive pulse formation is not possible in QCL combs even in the case of zero dispersion, and that their output resembles the one of continueous wave lasers~\cite{Khurgin:2014hy,Villares:2015ho}. Theoretically, the phase relation of the individual comb modes is fixed, but can have more realizations than the one required for Fourier-limited pulses. In the time-domain, this can result in full frequency modulation or a mixture between amplitude and frequency modulation. Experimentally, the only successful approaches to generate pulses from QCLs are indirect, based on injection seeding the active medium with picosecond short pulses from a photoconductive antenna~\cite{Bachmann:16,Soibel:2004p1534}, or on active mode-locking~\cite{Barbieri:2011p1986}. In these particular cases, the emitted terahertz light is phase-locked to a reference femtosecond oscillator, and its output profile could be measured by coherent electro-optic sampling~\cite{Kroell:2007p211,Oustinov2010}. For the case of free-running QC combs, coherent electr-optic sampling is not feasible and the method of choice for the characterization of their peculiar properties has long been beat-note analysis on a sufficiently fast detector, in the absence of intensity autocorrelation techniques at Terahertz frequencies~\cite{Burghoff:2015jf}. 

In conventional NIR frequency combs, the coherence properties are routinely assessed by intensity autocorrelation, which is based on second harmonic generation of the lightwave with a delayed replica using a $\chi^{(2)}$ medium~\cite{Armstrong1967,Hirayama2002}.  A simmilar technique would be of great help for the characterization of combs in the Terahertz range, however, any nonlinear scheme is suited in particular for the characterization of high peak power pulsed radiation, since the signal is scaling like $I^2$. So far, Quantum cascade laser based frequency combs are expected to emmit continuously~\cite{Burghoff:2015jf}, and would have a weak non-linear signal because they are diluted in time. In addition, the THz frequency range has long been completely lacking such tool to measure intensity autocorrelations. Recently, a novel technique has been reported to measure electric field and intensity correlations at THz frequencies by electro-optic sampling with a  very short temporal resolution of 146~fs~\cite{Benea-Chelmus2016}. This technique is optimally suited for the investigation of long- and short-scale correlations of QCL based frequency combs, since the temporal resolution of this technique is high enough to resolve temporal dynamics as fast as the gain recovery time, typically only few picoseconds~\cite{Choi:2008p303}. In addition, even cavity roundtrip times of such lasers are as short as few tens of picoseconds, and correspond to cavity lengths of few millimiters. 

In the present work, we exploit autocorrelation to get insight into the ouput dynamics of THz QCL frequency combs. Several devices have been tested and show qualitatively similar behaviour. From them, one will be presented in the paper and another one can be found in the supplementary material. The laser is operating as a low-noise frequency comb at low input current where it spans the spectral range from 2.2-2.8~THz, and in the dispersed regime at high input current, where the emission is more broadband, reaching a maximum from 1.6-3.3~THz. The detailed performance and spectral characteristics of the investigated device have been reported in reference~\cite{Rosch:2014ft,Faist2016}. We compare the results with the emission profile of a  pulsed frequency comb in the Terahertz based on optical rectification in a photoconductive antenna.

A schematics of the correlation measurement method is shown in figure~\ref{fig:setup}~(a). The operation principle relies on the non-linear interaction~(Pockels) between the THz wave and two mutually delayed~($\tau$) ultra-short near-infrared~(NIR) probing pulses~\cite{Benea-Chelmus2016}. They are in essence sampling the electric field of the THz wave at two distinct points in time~($t_i$), with the associated measurement results denoted in the following $\mathcal{E}(t_i)$ and $\mathcal{E}(t_i+\tau)$. A fast two-channel analog-to-digital converter is utilized to record in real-time the electric field value of each pair of pulses, $\mathcal{E}(t_i)$ and $\mathcal{E}(t_i+\tau)$, at the speed of the repetition rate of the laser, in our case 90~MHz.  From this single measurement, and subsequent averaging over the result of such single measurements, the instantaneous intensity is retrieved, as well as electric field and intensity autocorrelations. The instantaneous intensity of the THz light wave is retrieved digitally, by applying the intensity formula  $\mathcal{I} = c\epsilon_0\epsilon_r\mathcal{E}^2$ on the corresponding electric field values. The electric field and intensity correlations are time averages of products as $\mathcal{E}_{THz}(t_i) \mathcal{E}_{THz}(t_i+\tau)$ and $\mathcal{I}_{THz}(t_i) \mathcal{I}_{THz}(t_i+\tau)$. They are retrieved by multiplying each pair of measurements and averaging over a big amount of sample pairs, which in the case of a free running source is equivalent to time-averaging over laboratory time-scales. Finally, we normalize these products are retieve the normalized autocorrelation functions, denoted $g^{(1)}(\tau)$ and $g^{(2)}(\tau)$ and  defined as follows. 
\begin{equation}
g^{(1)} (\tau)= \frac{ \langle \mathcal{E}_{THz}(t) \mathcal{E}_{THz}(t+\tau) \rangle_{t} }{\sqrt{\langle \mathcal{E}_{THz}(t)  \mathcal{E}_{THz}(t)  \rangle_{t} \langle \mathcal{E}_{THz}(t+\tau)  \mathcal{E}_{THz}(t+\tau) \rangle_{t}}}
\end{equation}

\begin{equation}
g^{(2)}(\tau) = \frac{ \langle \mathcal{E}_{THz}(t)\mathcal{E}_{THz}(t+\tau) \mathcal{E}_{THz}(t)\mathcal{E}_{THz}(t+\tau) \rangle_{t} }{\langle  \mathcal{E}_{THz}(t)  \mathcal{E}_{THz}(t) \rangle_{t} \langle \mathcal{E}_{THz}(t+\tau)  \mathcal{E}_{THz}(t+\tau) \rangle_{t}}.
\end{equation}

and $\langle~\rangle_{t}$ represent averages over time. 

 A detailed description of the setup and measurement algorithm can be found in reference~\cite{Benea-Chelmus2016}. 

Each of these quantities will be utilized to infer complementary information about the operation mode of the comb lasers. The first order coherence function  $g^{(1)}(\tau)$ will give us, similarily like a Fourier Transform Infrared Spectrometer the power spectrum of emission, through the Wiener Khinchin theorem~\cite{Vetterli2014}. The frequency resolution is determined by the temporal delay, which in our case is roughly 60~ps, corresponding to 17~GHz. The value at $\tau=0$ is $g^{(1)}(\tau = 0)=1$, as is evident from the definition. The second order coherence function  $g^{(2)}(\tau)$ will be used to infer information about the output pattern of the laser, since it represents the normalized time-dependent autocorrelation of the intensity. The quantity  $g^{(2)}(\tau)$  is 1 when the intensity $I(t)$ and $I(t+\tau)$ are fully uncorrelated and can be factorized:
\begin{equation}
g^{(2)}(\tau) = \frac{\langle I(t)I(t+\tau) \rangle_t}{\langle I(t) \rangle_t \langle I(t+\tau) \rangle_t} =  \frac{\langle I(t)\rangle_t\langle I(t+\tau) \rangle_t}{\langle I(t) \rangle_t \langle I(t+\tau) \rangle_t}  =1
\end{equation}

More importantly for our analysis about the output profile of comb lasers, for classical radiation, the value of $g^{(2)}(\tau=0)$ gives particular insight into the intensity modulation depth, since it can be rewritten as:

\begin{equation}
g^{(2)}(\tau=0) = \frac{\langle I(t)^2 \rangle_t}{\langle I(t) \rangle_t^2} = 1+\frac{ Var(I(t)) }{\langle I(t) \rangle_t^2} 
\end{equation}
where $ Var(I(t))$ represents the variance of the intensity.  From here, we define the amplitude modulation depth in the following way:
\begin{equation}
\frac{\sqrt{Var(I(t))}}{\langle I(t) \rangle_t}= \sqrt{g^{(2)}(\tau=0)-1}
\end{equation}

\begin{figure*}[hbtp]
  \centering 
  \includegraphics[width=\textwidth]{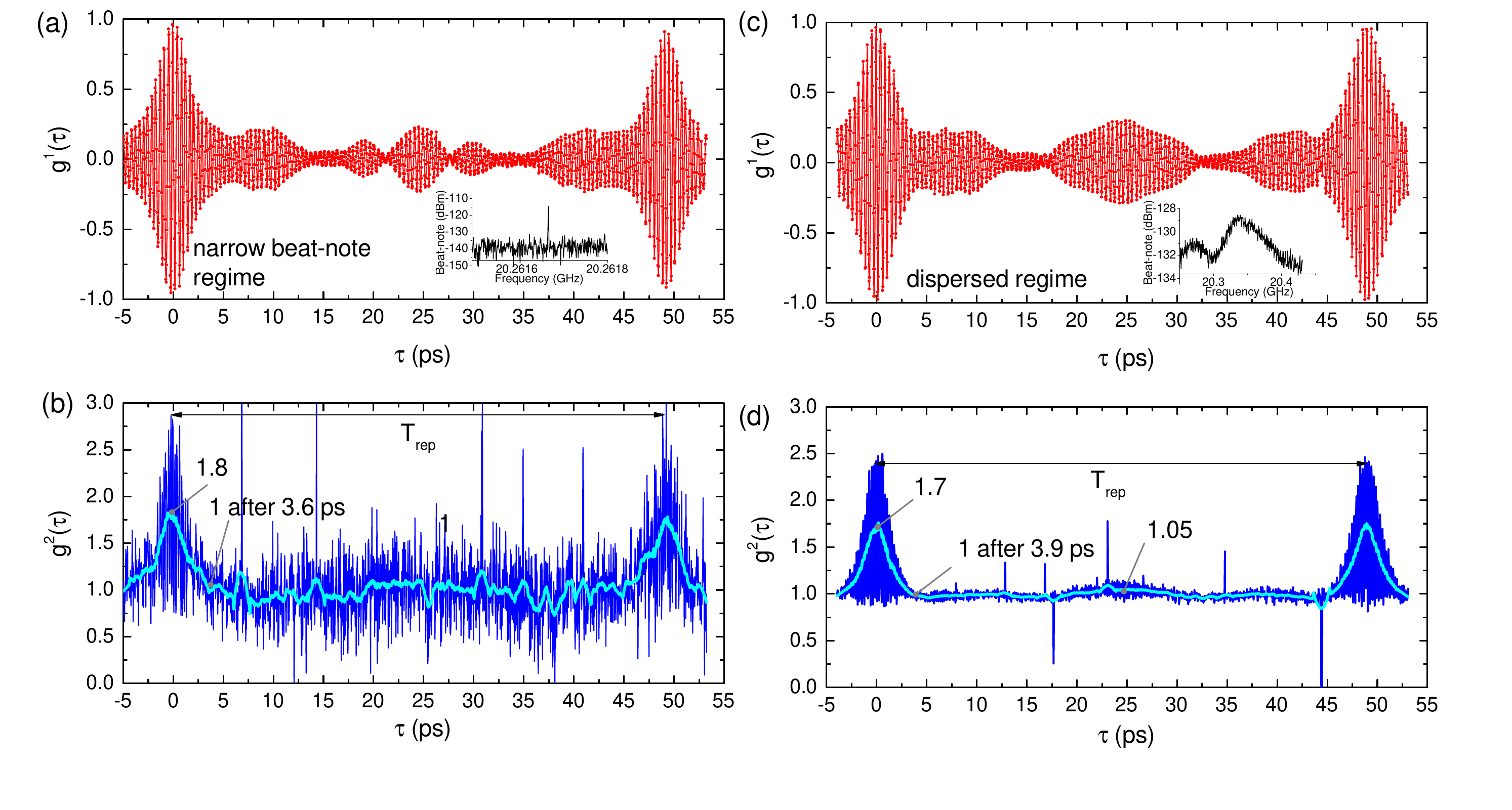}
  \caption{First and second order coherence measurements of a 2~mm long THz laser comb operating in the narrow beat note regime~((a)-(b)) and the dispersed regime~((c)-(d)). (a)~$g^{(1)}(\tau)$  measurement of laser emission in the narrow beat-note regime, at 906~mA and T=23~K. Shown in the inset is the beat-note as measured when operating the laser in continuous wave at 900~mA and the beat-note is as short as 800~Hz. (b) $g^{(2)}(\tau)$  in the narrow beat-note regime. For long time delays, the intensity is uncorrelated, and decays to a value of 1, typical for a continuous-wave like output. At $\tau = 0$,   $g^{(2)}(0)$ = 1.8, and therefore the intensity modulation depth $\frac{ \sqrt{Var(I(t))} }{\langle I(t) \rangle_t} =0.9$. (c) $g^{(1)}(\tau)$ measurement of laser emission in the dispersed regime, at 1010~mA and T=23~K.  Shown in the inset is the beat-note as measured when operating the laser in continuous wave at 1000~mA. (d) $g^{(2)}(\tau)$ measurement of laser emission in the dispersed regime. For long time delays, the intensity becomes uncorrelated, and decays again to a value of 1. Also in the dispersed regime, the emission has a continuous-wave like output. At $\tau = 0$,   $g^{(2)}(0)$ = 1.7,  and therefore the intensity modulation depth $\frac{ \sqrt{Var(I(t))} }{\langle I(t) \rangle_t} =0.83$. The time delay for both measurements is longer than the cavity round-trip time which in this case is 49.5~ps, corresponding to a 2~mm long, 150~$\mu$m wide double metal ridge. Light blue lines represent few-cycle average of the  $g^{(2)}(\tau)$, to cancel the sub-cyle resolution of the intensity autocorrelation~(window averaging 1.66~ps).}
  \label{fig:g1g2togeter}
\end{figure*}

The double beam technique is self-referenced. In comparison to coherent sampling, it brings the advantage that free-running sources can be investigated herewith. Also, even incoherent radiation which is otherwise averaged out by coherent sampling techniques, is measured together with the coherent part of the signal. All these essential properties make this technique perfectly suited for the investigation of temporal dynamics in THz comb lasers. 

For enhanced sensitivity, we match the emission bandwidth of the laser with the detection bandwidth of the nonlinear effect by choosing  a 3~mm long Zinc Telluride~(ZnTe, (110)) crystal as detection medium. In this configuration, it exhibits a coherence window in the spectral region around 2.4~THz where the laser is operating, if combined with a probe central wavelength of 775~nm. In addition, the detection bandwidth is several hundrets of GHz. Figure~\ref{fig:setup}~b represents the computed coherence length in this configuration, using $l_c=\frac{\pi c}{2\pi f_{THz} |n_{g,opt}-n_{ph,THz}|}$, with marked spectral regions where the laser is emitting in the comb~(2.2-2.8~THz) and dispersed regime~(1.6-3.3~THz). 

\begin{figure*}[hbtp]
  \centering 
  \includegraphics[width=\textwidth]{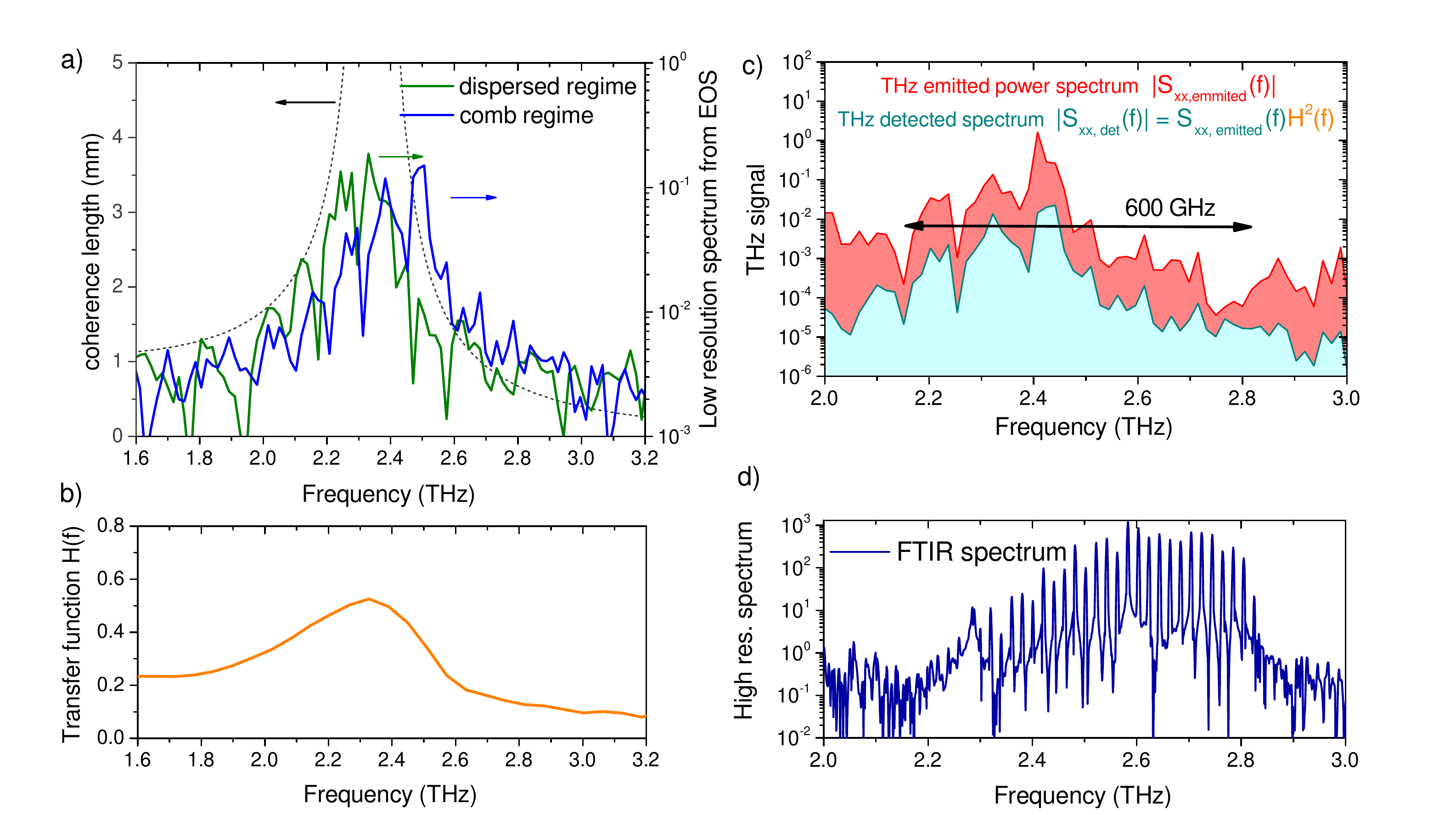}
  \caption{Spectral analysis of the QCL based frequency comb when operated in the narrow beat-note and dispersed regime, using electro-optic correlations. (a) Low resolution spectra from the Fourier Transformation of $g^{(1)}(\tau)$ (green and blue), shown together with coherence length computations for ZnTe and a center probe wavelength of 775~nm. (b) Transfer function of the detection based on ZnTe under these conditions, at room temperature. (c) Raw THz power spectrum and deconvoluted emission power spectrum using the transfer function shown in (b). (d) For comparison, the high resolution Fourier Transform Infrared Spectroscopy power spectrum taken at 900~mA coincides well with the deconvoluted spectrum by electro-optic sampling~\cite{Rosch:2014ft,Faist2016}.}
  \label{fig:spectra}
\end{figure*}

\section{Current-dependent correlations}

We were interested in the dynamics of the QCL based laser comb when operated in the narrow-beatnote regime, at 906~mA, and dispersed regime, at 1010~mA. The QCL based frequency comb was operated in quasi continuous-wave~(CW) mode with a very slow modulation frequency of 2.5~kHz. The CW output power of the laser at these two operating points was 1.05~mW and 2.96~mW, respectively, as measured with a Thomas Keithling power meter.

The laser is 2~mm long and 150~$\mu$m wide, corresponding to a repetition rate of 20.3~GHz. The cavity of the laser is constituted of a double metal cavity for good temperature and dispersion performance and cryocooled to 23~K. The beat-notes of the comb lasers were measured when operated in continuous wave. 

The measured first and second order coherence functions are reported in figure~\ref{fig:g1g2togeter} as a function of time delay $\tau$, which covers more than the full round-trip time of the laser cavity, 49.5~ps. The electric field interferogram $g^{(1)}(\tau)$, is reported at the two operation points in figure~\ref{fig:g1g2togeter}(a)\&(c), and exhibits a repetition period of 49.5~ps, corresponding to the expected repetition rate of 20.3~GHz. At $\tau = 0$, $g^{(1)}(\tau)=0.98$ at both operating points, approximating well the theoretically expected value of 1. 

The intensity autocorrelation  $g^{(2)}(\tau)$ is reported in figure~\ref{fig:g1g2togeter}(b)\&(d). As elaborated in the introduction, the rather unusual property of electro-optic correlations is the temporal resolution, meaning that we resolve the instantaneous intensity of the laser. The intensity autocorrelation is therefore also periodic, with a characteristic frequency of double the emission frequency. In principle, correlated intensity fluctuations occuring at sub-cycle timescales can be seen in this way. To retrieve long-range behaviour,  we performed a running average filtering~(window width of 1.66~ps) to filter out these oscillations, and the result is reported by the cyan-colored line. In the comb regime, the intensity autocorrelation function takes  a value of 1.8 at $\tau=0$. Using the above formula, this suggests that $\frac{ \sqrt{Var(I(t))} }{\langle I(t) \rangle_t} = \sqrt{0.8} = 0.9$, and consequently, that the intensity modulation depth as defined presviously is roughly 90\%. In addition, $g^{(2)}(\tau)$ decreases to a value of 1 after roughly 3.6~ps, showing that correlated intensity modulations take place on timescales as short as 3.9~ps. At longer time-scales, the intensity autocorrelation function is roughly constant to 1, which suggests the peculiar continuous wave-like output. Clearly, in this case, even if the field is coherent, suggested by the interference fringes, the intensity itself has no predetermined correlations. These characteristics can therefore be sumed-up to the conclusion that the laser is both frequency and amplitude modulated. In the dispersed regime, $g^{(2)}(\tau)$ decreases to 1 after 3.9~ps, and decays from a value of 1.7 at $\tau=0$. Similarly, this suggests that $\frac{\sqrt{Var(I(t))}}{\langle I(t) \rangle_t} = \sqrt{0.7} = 0.83$, and consequently, that the intensity modulation depth is roughly 83\%. In addition, this intensity modulation takes place on longer timescales, 3.9~ps. Also in the dispersed regime, the laser is both frequency and amplitude modulated, with slightly smaller modulation depth on faster timescales. 

We investigated a second laser, 2.2~mm long, which was also operating in the comb and dispersed regime. We found qualitatively very simmilar results, namely a continuous wave like output with intensity correlations on fast timescales. This stands in high contrast to results that are to be obtained from pulsed sources, which we will discuss in the last section of this paper. The results are presented in the supplementary material.

\begin{figure*}[ht]
  \centering 
  \includegraphics[width=\textwidth]{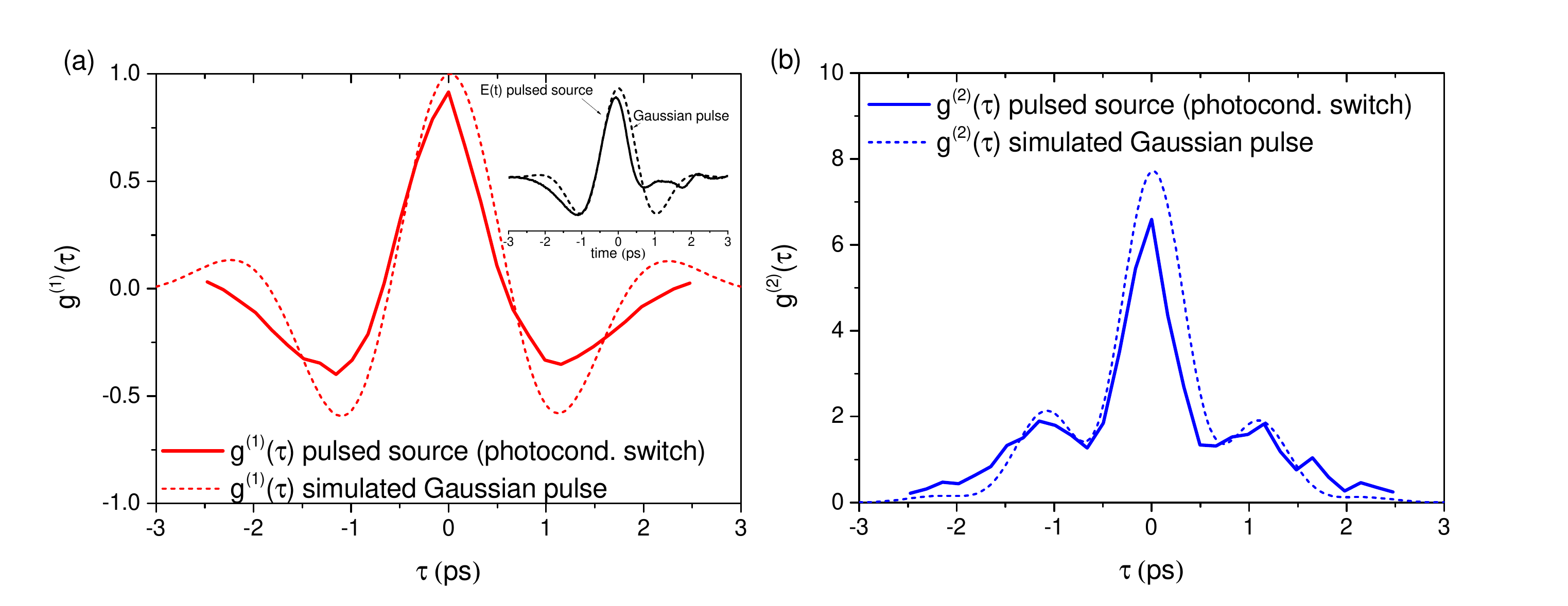}
  \caption{Results of first and second order coherence of a pulsed THz comb based on optical rectifictation in a photoconductive emitter upon excitation with femtosecond laser pulses. (a) $g^{(1)}(\tau)$ of the THz pulse  with an electric field profile as shown in the inset. We compare the measurement to a simulated pulse with Gaussian spectrum and find qualitatively good agreement. (b)  $g^{(2)}(\tau)$ of the pulsed radiation, normalized to the average intensity within a time window of 13~ps. Characteristic to pulsed radiation is the decay of the intensity autocorrelation function to a value of 0, which is a clear indicator for the pulsed emission. Additionally, the value at zero path delay of 6.59, corresponding to an intensity modulation depth of 242\%.}
  \label{fig:g1g2_PCA}
\end{figure*}

\section{Bandwidth of detection}

Our technique is spectrally resolving, and gives access also to the spectral information. The direct measurement of electric field auto-correlation $g^{(1)}(\tau)$ has been Fourier-transformed to retrieve the low resolution spectrum of the THz source. The results are reported in figure~\ref{fig:spectra}~a, for the comb and dispersed regime, together with the coherence properties of ZnTe. The emission spectra are significantly different in the two regimes. The low-current spectra reveals a spectral bandwidth of roughly 600~GHz centered around 2.6~THz. However, the spectral response function of ZnTe, here denoted $H(f)$, is not flat over the full bandwidth of interest, and therefore, the measured signal $E_{detected}$ is a convolution of the real laser emission $E_{emitted}(f)$ and the detection sensitivity of the ZnTe crystal:
$$E_{detected}(f) = E_{emitted}H(f)$$

The transfer function of a 3~mm long ZnTe crystal is reported in figure~\ref{fig:spectra}b, as measured with standard time domain spectroscopy from comparison with a very thin crystal~(200~$\mu$m), where a flat response was assumed. It includes both the effects of coherence and of THz absorption in ZnTe, which is particularily strong in this region due to phonon resonances. The transfer function is further used to retrieve the power spectrum of the THz source, as reported in figure~\ref{fig:spectra}c:

$$S_{xx,detected}(f) = S_{xx,emitted}H^2(f)$$

Finally, we compare the low resolution spectra as obtained by electro-optic sampling with the high-resolution Fourier Transform Infrared Spectroscopy setup. We find qualitatively good agreement.

\section{Intensity correlations of pulsed light}

We have demonstrated so far, that THz quantum cascade laser based frequency combs are both amplitude and frequency modulated, and have an output simmilar to continuous wave sources. This peculiar properties remain in big contrast to conventional frequency combs, where the emission is typically pulsed. Such pulsed sources have been realized in the Terahertz domain by optical rectification of femtosecond pulses in $\chi^{(2)}$ media. 

In the following, we will show that intensity autocorrelation results from pulsed sources are quantitatively very different from results obtained from quantum cascade laser frequency combs. With this goal, we measure the field and intensity autocorrelation functions of the pulsed THz radiation emitted by a photoconductive emitter, by optical rectification of femtosecond laser pulses at 775~nm. The results are shown in figure~\ref{fig:g1g2_PCA}, together with analytical simulations, which assume a simplified scenario of a THz pulse with Gaussian spectrum, no dispersion and a spectral width roughly coinciding with the one of the photoconductive emitter. This simplified example is instructive enough to reproduce the main characteristics. 

The electric field of the emission has been measured by coherent electro-optic sampling and is shown in the inset in figure~\ref{fig:g1g2_PCA}a. The autocorrelation functions have been measured by the two beam technique utilized also for the laser comb. They are measured for a time delay as long as the main temporal extent of pulse, roughly 6~ps. The first order coherence function $g^{(1)}(\tau)$ has a symmetric shape around $\tau = 0$, as expected. The most instructive characteristic is however in the shape and peak value of the second order coherence function  $g^{(2)}(\tau)$. It decays to a value of 0 after roughly 2.5~ps, which suggests the pulsed character of the emission. This is in strong contrast to the emission of quantum cascade laser combs, where the continuous wave-like output results in a $g^{(2)}(\tau)$ which decays to 1. Moreover, $g^{(2)}(0)$ at zero path delay is 6.59, when normalized to the average intensity in a time window of 13~ps. From this, we conclude, that in this time window, the intensity modulation depth is 242\%, considerably bigger than the value retrieved for mixed frequency and amplitude modulated combs. For completeness, these results can be also compared to measurements of single mode lasers, where $g^{(2)}(0)$ is 1, and therefore the intensity modulation depth is 0\%. Such results can be found in reference~\cite{Benea-Chelmus2016}.

\section{Discussion and conclusion}
We have demonstrated the first direct measurement of intensity correlations of a quantum cascade laser based frequency comb, and compared these results to mode-locked THz combs generated by optical rectification in a photoconductive emitter. We found that the laser is both amplitude and frequency modulated with a modulation depth of 90\% in the comb regime and 83\% in the dispersed regime. Moreover, the continuous-wave like output of the laser shows up in the intensity autocorrelation which decays to 1, in contrast to pulsed sources where it decays to 0.

For this purpose, we utilized a technique which has a temporal resolution of 146~fs, only a fraction of one period of light, and is in addition sensitive to any incoherent components in the lightwave through its selfreferencing character. All these characteristics render the technique unique in its features. 

Having the possibility to fastly assess the emission profile of free-running lasers, is an important technological milestone. In future, inherently frequency modulated combs based on quantum cascade lasers, could be transformed into amplitude modulated combs by use of an exernal spatial light modulator which is dynamically shifting the phases which are stable, but do not fulfill the mutual condition for Fourier limited pulses. Our technique will be here of big importance to minimize in real-time the temporal extent and provide feedback to the actuating modulator. 

More generally, the technique is perfectly suited to measure flutuations faster than one oscillation period, thus sub-cycle. This allows for many more fundamental experiments to be envisaged.

\section*{Funding Information}

The authors would like to acknowledge funding from the  ERC (Advanced Grant, ‘Quantum Metamaterials in the Ultra Strong Coupling Regime’). 

\section*{Acknowledgments}

We acknowledge the insightful comments of Christopher Bonzon on the data and his expertise with cryogenic measurements. We also acknowledge the work of the mechanical workshop at ETHZ. The sample growth and processing took place in the clean room facility of ETHZ, FIRST center.

\section*{Supplemental Documents}

\bigskip \noindent See \href{link}{Supplement 1} for supporting content.

\end{document}